\documentclass{article}
\usepackage{spconf,amsmath,graphicx}
\usepackage{hyperref}
\usepackage{svg}
\usepackage{lipsum}
\usepackage{multirow}
\usepackage{dcolumn}
\usepackage{array}
\usepackage{booktabs}
\DeclareMathVersion{nxbold}
\SetSymbolFont{operators}{nxbold}{OT1}{cmr} {b}{n}
\SetSymbolFont{letters}  {nxbold}{OML}{cmm} {b}{it}
\SetSymbolFont{symbols}  {nxbold}{OMS}{cmsy}{b}{n}

\title{J-Net: Randomly weighted U-Net for audio source separation}
\name{Bo-Wen Chen$^{\dagger}$ \qquad Yen-Min Hsu$^{\star}$ \qquad Hung-Yi Lee$^{\dagger \star}$}
\address{$^{\star}$Department of Electrical Engineering, National Taiwan University\\
	     $^{\dagger}$Graduate Institute of Communication Engineering, National Taiwan University}
\begin{document}
\ninept

\maketitle
\begin{abstract}

Several results in the computer vision literature have shown the potential of randomly weighted neural networks. While they perform fairly well as feature extractors for discriminative tasks, a positive correlation exists between their performance and their fully trained counterparts. According to these discoveries, we pose two questions: what is the value of randomly weighted networks in difficult generative audio tasks such as audio source separation and does such positive correlation still exist when it comes to large random networks and their trained counterparts? 


In this paper, we demonstrate that the positive correlation still exists. Based on this discovery, we can try out different architecture designs or tricks without training the whole model. Meanwhile, we find a surprising result that in comparison to the non-trained encoder (down-sample path) in Wave-U-Net, fixing the decoder (up-sample path) to random weights results in better performance, almost comparable to the fully trained model.



\end{abstract}
\section{Introduction}
\label{sec:intro}
Some works in computer vision \cite{1,2,3,16} or audio processing \cite{4} have shown that randomly weighted networks have the potential to extract features for discriminative tasks like classification. In particular, Saxe et al. \cite{1} found that the performances of randomly weighted networks reveal the intrinsic potentials and characteristics of their architecture. Besides, the performance of randomly weighted models is strongly correlated with the performance of their fully trained counterparts under the same architectures. Thus, we can conduct an architecture search without a time-consuming procedure to train the whole model. Rosenfeld and Tsotsos \cite{2} demonstrated a negligible decrease in the performance of image classification when models fix most of their weights to random. Their results also strengthen the argument of Saxe et al. \cite{1} that a strong correlation between trained and non-trained models exists within the performance.

Few related works about random models exist in the field of audio processing. Pons and Serra \cite{4} extracts features from randomly weighted CNN and inputs those features to support vector machines (SVM) or extreme learning machines (ELM) with the goal to compare classification accuracies when using different randomly weighted architectures.  They also compare features from different architectures with MFCCs as a baseline. Some of their results are even comparable to trained CNNs, revealing that the architecture alone imposes many contributions to the audio prior. Besides, they find that sample-level features are better than frame-level which is also concluded in \cite{17,18}. However, they didn't compare the randomly weighted CNN to the corresponding trained model.

Michelashvili and Wolf \cite{15}, inspired by Dmitry Ulyanov \cite{16}, developed an unsupervised method for denoising, which utilized the audio prior reserved in the architecture of Wave-U-Net \cite{5}. They find that different from image, there is little difference between the time taken to fit noise added audio and to fit clean audio. Thus, they first transform the output of Wave-U-Net into STFT and multiply it by a mask which subtracts the part of STFT with less stability.

Wave-U-Net \cite{5} is currently the state-of-the-art waveform based model for audio source separation, variant of U-Net \cite{6} architecture adapted for the one-dimensional time domain. While the performance is good, the long training time is quite a pain, making architecture search on variants of Wave-U-Net difficult To our best knowledge, we are the first to try a randomly weighted model on audio source separation, a difficult generative task in comparison to other discriminative tasks.

Our work aims to validate the positive correlation between performances of randomly weighted model with the trained counterparts. To this end, in Section 2, we present several method adaptations proposed by other work which are shown effective for different tasks in computer vision, creating several architectures with Wave-U-Net as the backbone. And we want to see whether the performance relation between these random weighted models with their trained counterparts confronts this argument. In Section 3 and 4, we evaluate the above-mentioned models, to conclude in Section 5 that not just the strong performance correlation exists, we also discover a counter-intuitive fact that Wave-U-Net with random weighted decoder has better performance in comparison to the counterparts with untrained encoder. Overall, our contributions in this paper can be summarised as follows.


\begin{itemize}
    \item We propose a framework for architecture search by evaluating the performance of model variants with untrained encoders.
    \item We discover some architecture settings which outperform the baseline model from Wave-U-Net.
    \item We discover U-Net with untrained decoder path has great potential which is almost comparable to fully trained models.
\end{itemize}

\section{METHODOLOGY}

\begin{figure*}[htb]
  \centering
  \centerline{\includegraphics[width=\linewidth]{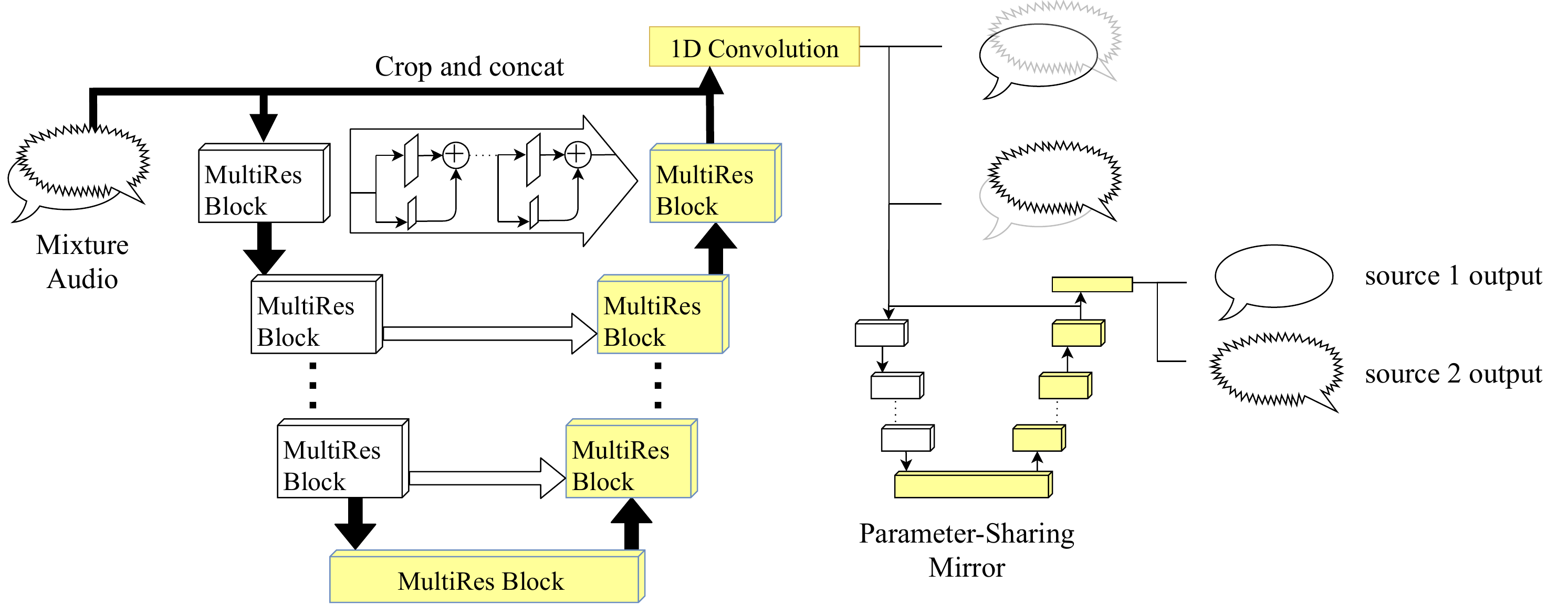}}

\caption{The approaches mentioned in Section 2. The arrow represents \textit{Res path} from Section 2.1. And for Section 2.2, we replace convolution layers in Wave-U-Net with \textit{MultiRes} block. The parameter-sharing mirror of J-Net is used by progressive learning from Section 2.3}
\label{fig:arch}
\end{figure*}

The base architecture is adapted from Wave-U-Net \cite{5}, an adaption of U-Net \cite{6} to the one-dimension time domain. Wave-U-Net is also in the form of encoder-decoder architecture with encoder successive down-sampling the feature maps, and decoder up-sampling the feature-map to reconstruct the target waveform. Also, the skip connections between encoder and decoder allow the decoder to further utilize the feature maps from the encoder. To validate the existence of correlation, we show several methods below to form different training and architectural criterion.

\subsection{Residual Path}
In \cite{7}, they argue that despite the main contribution of U-Net came from the introduction of skip connections, there may be probable semantic gap between the corresponding levels of encoder and decoder. The features from encoder are supposed to be lower level, while the features from decoder are at a much higher level, the conjecture of such incompatible features might worsen the final generalization. To tackle this problem, they propose \textit{Res path}, which replace direct skip connection with several CNN layers and non-linear transformations to reconcile these two incompatible sets of features from encoder and decoder. We add the \textit{Res path} proposed in \cite{7}, shown as arrow in Fig. 1, to our model to examine whether 'Res path' can solve the semantic gap problem and further increase the performance of origin Wave-U-Net.

\subsection{MultiRes Block}
In \cite{7}, they mention the problem of variation of scale in biomedical images. Therefore, networks are supposed to be robust to analyze objects of interest at irregular scales. To solve this problem, they replace the successive two 3x3 convolution layers of original U-Net \cite{6} in both encoder and decoder with \textit{MultiRes} block, which is composed of few 3x3 convolution layers with each layer's output concatenated together as final block output. Since successive small filters have the same effect as one large filter \cite{19}, 'MultiRes' block can perform multi-resolution analysis. In the original paper, the number of filters in the corresponding layer in \textit{MultiRes} block is assigned by a criterion to control the number of parameters not to exceed initial U-Net \cite{6}. However, in this paper, we replace one or two successive convolution layers to a \textit{MultiRes} block, as shown in Fig. 1, containing two convolution layers with same number of filter and same filter size as corresponding replaced layers.

\subsection{Progressive Learning}
There are several works on progressive learning (PL), both in computer vision \cite{10,13} and speech enhancement \cite{11,12}. While drastically reducing the amount of parameters, output quality is also greatly improved. We propose an architecture which is composed of two stages of Wave-U-Net, with the parameters shared between corresponding layers between two Wave-U-Nets. In other words, we pass the mix audio into Wave-U-Net, for the first time, and then pass each separated sources into Wave-U-Net again to get the fully separated outputs as shown in Fig.1. The output of these two stages are both guided by the same clean source audios respectively. Also, we test the separation results with more than one stage by the baseline model to verify the actual benefits gained from progressive learning.

\subsection{Identity Loss}
In \cite{8,9}, they use identity loss to better preserve the content from the original domain, and to enhance the robustness of model by preventing mapping images from target domain to different images. 

Since our task can also be viewed as an audio translation task from mix domain $\mathbf{S}$ to vocal domain $\mathbf{X_1}$ or accompaniment domain $\mathbf{X_2}$, we adapt identity loss to enhance the robustness of our model. We map source $x \in \mathbf{X}$ in each source domain $\mathbf{X_i}$ to $f(x) \in \mathbf{X_i}$ within the same corresponding domain $\mathbf{X_i}$, where $f:S \rightarrow X_i$ represents the corresponding output of Wave-U-Net. Our identity loss $\mathcal{L}_{idt}$ can be formulated as:
\begin{equation*} 
\mathcal{L}_{idt} = \sum_{x\in \mathbf{X}_i}{d_2(x,f(x))} 
\end{equation*} 
where $d_2$ denotes the L2 loss. 


\section{Experiments}
\label{sec:experiments}
\subsection{Datasets}
\label{ssec:datasets}

We evaluated our proposed method on the publicly available dataset \cite{22}, a multi-track database for separation. It's split into a training partition of 100 tracks and a test partition of 50 tracks. Our validation setting is kept the same as \cite{5}, where 75 tracks from the training partition are selected randomly to form our training set and the remain 25 tracks are for validation. Also, the whole CCMixter dataset \cite{24} is added to the training set. All tracks are sampled at 22050Hz, and retain stereo. The only data augmentation method we use is described in \cite{5}.

\subsection{Training procedure}
\label{ssec:training_proc}
Audio snippets are randomly sampled for training. Different from \cite{5}, the output of our every models have the same length as input, thus, there is no need to pad input. With respect to loss, we use the mean squared error (MSE) as loss source output samples, and average the MSE loss according to the batch size. As in \cite{5}, we use ADAM optimizer with settings recommended by Kingma and Ba in \cite{20}, which sets initial learning rate to 0.0001, decay rates $\beta_1$ to 0.9 and $\beta_2$ to 0.999. The initial batch size is 16, and each epoch contains 2000 iterations. We perform early stopping after 15 epochs of no improvement on the validation set with respect to MSE loss. After first early stopping occurs, we enter first fine-tune stage, where we multiply the batch size by two, and reset the learning to 0.00001. Again until 15 epochs without improvement on validation set, we enter final fine-tune stage, where learning rate is set to 0.000001 and batch unchanges. Lastly, the best model is selected depending on the validation loss.

\section{Variants of model settings}
\label{sec:model settings}
All our models can be grouped into two sets, one contains models with encoder fixed to random weights while the others are the corresponding fully trained counterparts with remaining architectural settings and training criterion staying the same. Since the shape of the model looks quite like alphabet "J" if only seeing the trained part, we call our model with randomly weighted encoder as "J-Net" while the trained counterpart as "U-Net". For our baseline model, we use samples of length 16384 for input and output. The number of convolution layers in both down-sampling and up-sampling path is set to 10. And the other settings considering amount of extra filters per layer and filter sizes remain the same as \cite{5} for simplicity.

To validate the argument that performance between randomly weighted model, J-Net and their trained counterpart, U-Net has strong positive correlation, we create lots of variant model settings by means mentioned in above Section 2. In addition, we also want to check whether those means have any positive effects or impacts over the performance of original Wave-U-Net. Our baseline model U1 is adapted from M4 from \cite{5} by removing the proper input context and re-sampling mentioned in Section 3.2.2 of \cite{5}, because the method forces the input and output to have different length, forming obstacles for methods in Section 2.3 and 2.4 to be applied.

J1 has the same model settings as U1, with the encoder untrained. U2\textsubscript{i\_j} adds the \textit{Res path} from Section 2.1 with i denoting the number of convolution layers reside in each \textit{Res path} and j as the number of skip connections. The convolution layer in \textit{Res path} has the same settings as the convolution layer at same level in encoder. Thus, dimension of embedding from each skip connection remains the same after applying \textit{Res path}. U3\textsubscript{10} replaces each layer in U1 with one \textit{MultiRes} block which contains two convolution layers with same number of filters and same filter size as the convolution layers which are replaced. In other words, there are 10 \textit{MultiRes} blocks in both down-sampling and up-sampling paths of U3\textsubscript{10}, while U3\textsubscript{5} contains only 5 \textit{MultiRes} blocks in each path. U4\textsubscript{i} adapts U1 with progressive learning method described in Section 2.3 and the index i denotes the number of stages for fully separation. And U5 add identity loss from Section 2.4 during training. The models which have gone through evaluation are U1, U2\textsubscript{2\_5}, U2\textsubscript{3\_5},  U2\textsubscript{2\_10},  U2\textsubscript{3\_10},  U3\textsubscript{5}, U3\textsubscript{10}, U4\textsubscript{1}, U4\textsubscript{2}, U4\textsubscript{3}, U5. Notice that U4\textsubscript{i} are same models but with i stage for separation.

The remaining model names which start with "J" represent untrained counterparts with only decoder/up-sampling path trained. For example, the \textit{Res path}es in J\textsubscript{i\_j} are also fix to random weights. Further implementation details can be found in our Github repository \url{https://github.com/EdwinYam/J-Net}.

Finally, in comparison to random encoder, we also evaluate the performance of random decoder. The model with all skip connections reserved is denoted as L since the remaining trained weights within the model have the shape like alphabet L. The model with skip connections only from the first three layers of encoder and model with skip connections kept only for the last three layers of encoder are denoted as L\textsubscript{first-}3 and L\textsubscript{last-3} respectively.

\section{Results}
\label{sec:results}

\begin{figure}[htb]

\begin{minipage}[b]{1.0\linewidth}
  \centering
  \centerline{\includegraphics[width=1.0\linewidth]{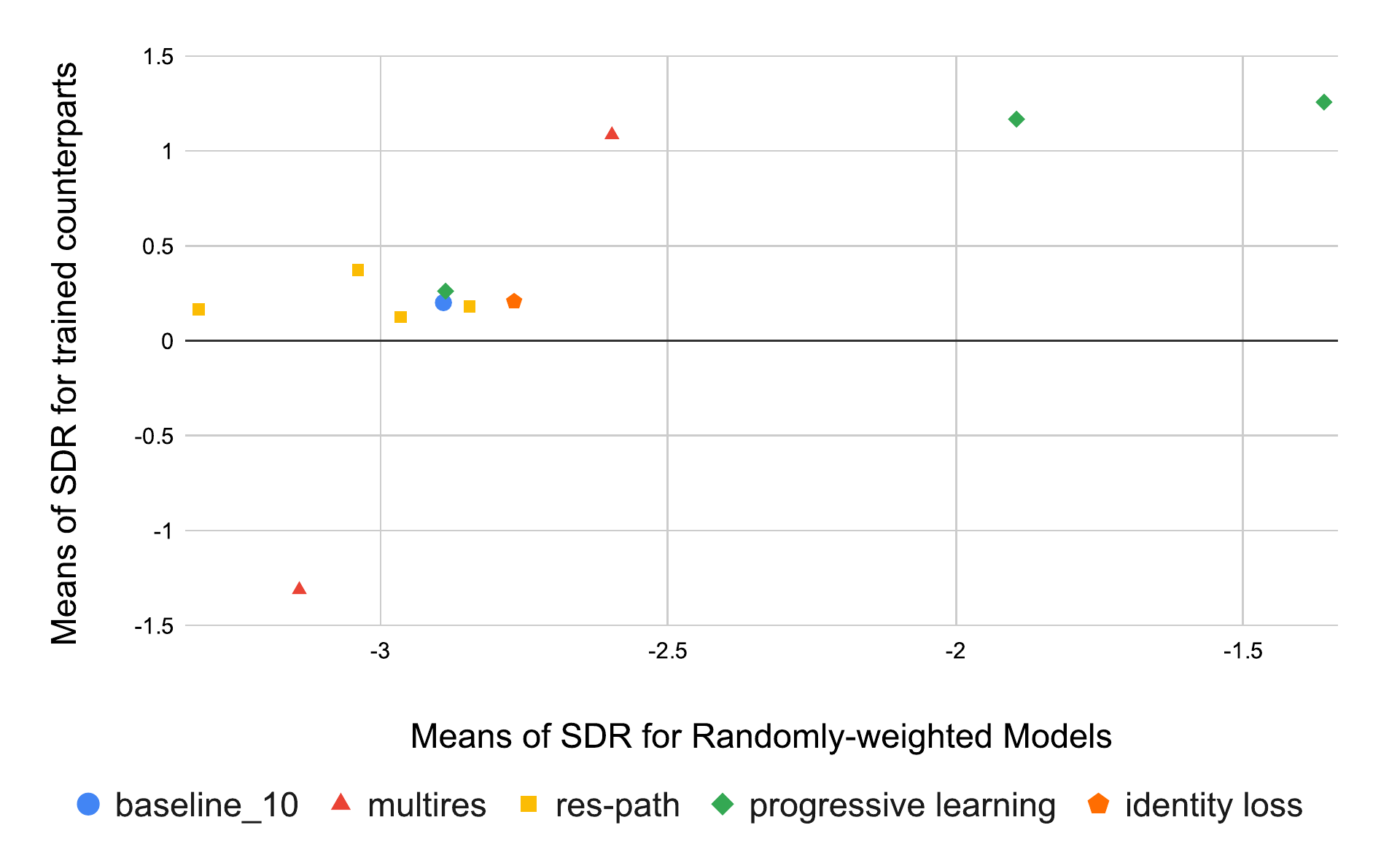}}

\end{minipage}
\caption{Above show the results of mean SDR for separated vocals. Taking the mean SDRs of vocals separated by U-Net, as fully trained model, as the x-axis and their counterparts by J-Net with encoder path remain untrained, as the y-axis}
\label{fig:res}

\vspace{1.0cm}
\begin{minipage}[b]{1.0\linewidth}
\centering
\centerline{\includegraphics[width=1.0\linewidth]{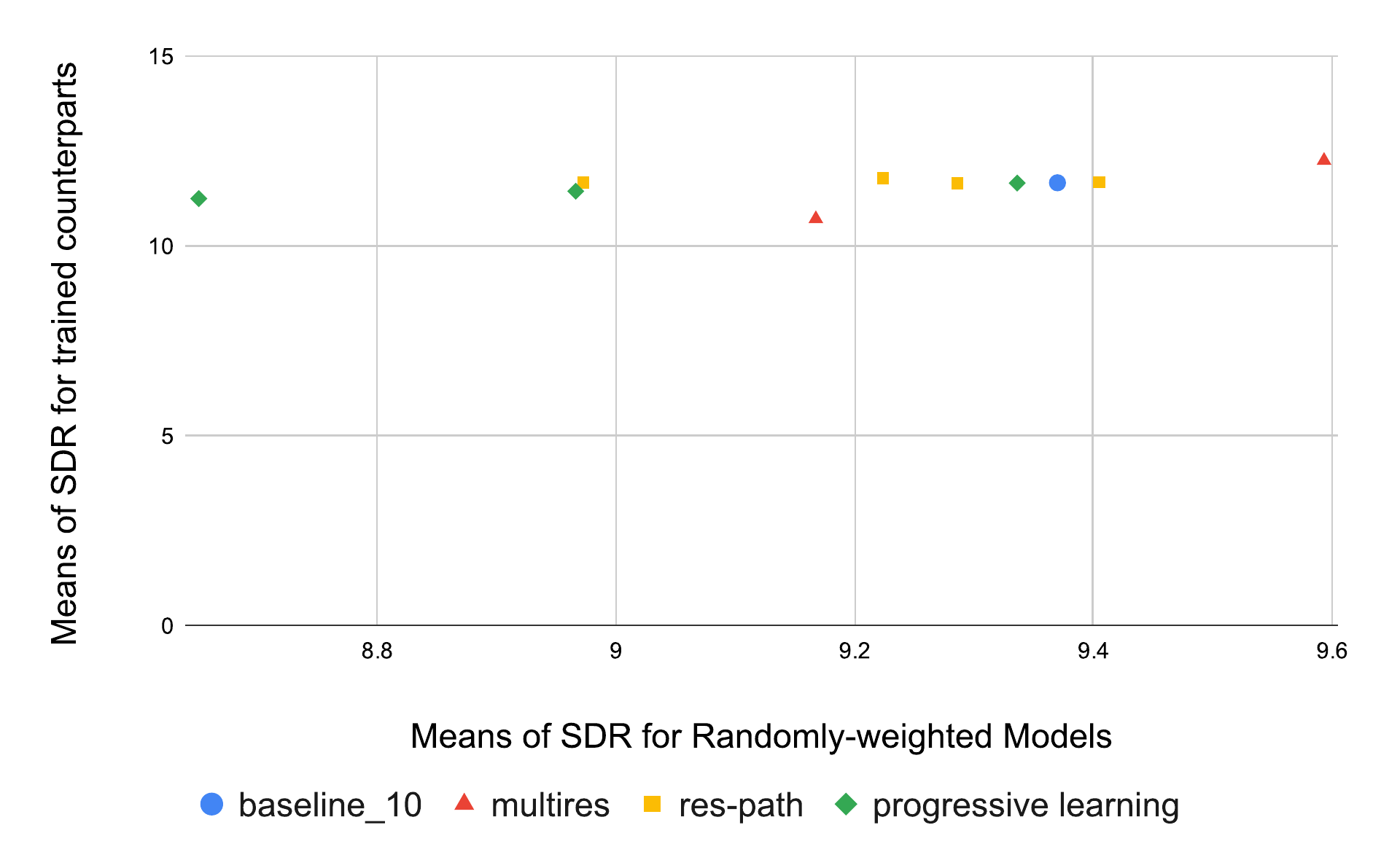}}
\end{minipage}
\caption{Above show the results of mean SDR for separated accompaniments. Taking the mean SDRs of vocals separated by U-Net, as fully trained model, as the x-axis and their counterparts by J-Net with encoder path remain untrained, as the y-axis}
\label{fig:res2}
\end{figure}

\makeatletter
\newcolumntype{B}[3]{>{\boldmath\DC@{#1}{#2}{#3}}c<{\DC@end}}
\newcolumntype{Z}[3]{>{\mathversion{nxbold}\DC@{#1}{#2}{#3}}c<{\DC@end}}
\makeatother
\begin{table}[!htb]
\scriptsize
\centering
\begin{tabular}{ |c c| D{.}{.}{2}D{.}{.}{2}D{.}{.}{2}D{.}{.}{2}D{.}{.}{2}D{.}{.}{2}| }
\hline
  & & \multicolumn{1}{r}{M4} & \multicolumn{1}{r}{U1} & \multicolumn{1}{r}{U3\textsubscript{10}} & \multicolumn{1}{r}{U2\textsubscript{2\_10}} & \multicolumn{1}{r}{U4\textsubscript{1}} & \multicolumn{1}{r|}{U5} \\ 
\hline
\multirow{4}{*}{Voc.} & Med. & 4.46 & 4.48 & \multicolumn{1}{Z{.}{.}{2}}{4.84} & \multicolumn{1}{Z{.}{.}{2}}{4.49} & 4.34 & 4.11 \\
& MAD &  3.21 & 3.25 & 3.33 & 3.28 & 3.26 & 3.19 \\  
& Mean & 0.65 & 0.20 & \multicolumn{1}{Z{.}{.}{2}}{1.09} & \multicolumn{1}{Z{.}{.}{2}}{0.38} & \multicolumn{1}{Z{.}{.}{2}}{0.26} & \multicolumn{1}{Z{.}{.}{2}|}{0.21} \\
& SD & 13.67 & 14.36 & 13.57 & 14.15 & 14.06 & 13.82 \\
\hline
\multirow{4}{*}{Acc.} & Med. & 10.69 & 10.43 & \multicolumn{1}{Z{.}{.}{2}}{10.91} & \multicolumn{1}{Z{.}{.}{2}}{10.61} & 10.33 & 10.05 \\
& MAD &  3.15 & 2.99 & 3.14 & 3.02 & 2.99 & 3.01 \\
& Mean & 11.85 & 11.68 & \multicolumn{1}{Z{.}{.}{2}}{12.26} & \multicolumn{1}{Z{.}{.}{2}}{11.80} & 11.67 & 11.47 \\
& SD &  7.03 & 6.65 & 6.84 & 6.59 & 6.68 & 6.92 \\
\hline
\end{tabular}
\caption {Performance of the models which outperform baseline U1 on top of mean SDR of output vocals. We also list the best result, M4 from \cite{5}}

\end{table}

\begin{table}[!htb]
\scriptsize
\centering
\begin{tabular}{ |c c| D{.}{.}{2}D{.}{.}{2}D{.}{.}{2}D{.}{.}{2}| }
\hline
  & & \multicolumn{1}{r}{U1} & \multicolumn{1}{r}{L} & \multicolumn{1}{r}{L\textsubscript{first-3}} & \multicolumn{1}{r|}{L\textsubscript{last-3}} \\ 
\hline
\multirow{4}{*}{Voc.} & Med. & 4.48 & 3.83 & 3.64 & 3.93 \\
& MAD & 3.25 & 2.91 & 2.84 & 2.91 \\  
& Mean & 0.20 & -1.39 & -1.49 & -1.07 \\
& SD & 14.36 & 15.47 & 15.39 & 15.19\\
\hline
\multirow{4}{*}{Acc.} & Med. & 10.43 & 9.71 & 9.66 & 9.91\\
& MAD & 2.99 & 2.78 & 2.86 & 2.88 \\
& Mean & 11.68 & 10.65 & 10.57 & 10.90 \\
& SD & 6.65 & 6.03 & 6.07 & 6.12 \\
\hline
\end{tabular}
\caption {Performance of the models with decoder fixed to random weights, comparing to the baseline model U1}
\end{table}
\vspace{1.0cm}
\begin{table}[!htb]
\scriptsize
\centering
\begin{tabular}{ |c c| D{.}{.}{2}D{.}{.}{2}D{.}{.}{2}D{.}{.}{2}D{.}{.}{2}D{.}{.}{2}| }
\hline
  & & \multicolumn{1}{r}{U4\textsubscript{1}} & \multicolumn{1}{r}{U4\textsubscript{2}} & \multicolumn{1}{r}{U4\textsubscript{3}} & \multicolumn{1}{r}{U1\textsubscript{1}} &
  \multicolumn{1}{r}{U1\textsubscript{2}} &
  \multicolumn{1}{r|}{U1\textsubscript{3}} \\ 
\hline
\multirow{4}{*}{Voc.} & Med. & 4.34 & 4.32 & 4.25 & 4.41 & 4.19 & 3.90 \\
& MAD & 3.26 & 3.24 & 3.16 & 3.25 & 3.17 & 3.03\\  
& Mean & 0.26 & 1.17 & 1.26 & 0.21 & 1.15 & 1.00\\
& SD & 14.06 & 12.49 & 12.06 & 14.21 & 12.05 & 11.55\\
\hline
\multirow{4}{*}{Acc.} & Med. & 10.33 & 10.30 & 10.27 & 10.44 & 10.38 & 10.34\\
& MAD & 2.99 & 2.97 & 2.95 & 3.03 & 2.98 & 2.96\\
& Mean & 11.67 & 11.46 & 11.26 & 11.69 & 11.44 & 11.20 \\
& SD & 6.68 & 6.38 & 6.13 & 6.86 & 6.55 & 6.31\\
\hline
\end{tabular}
\caption {The results of progressive learning on Wave-U-Net. U4\textsubscript{i} is the model applied with progressive learning, while U1 is the baseline model. The subscripts indicate the number of stages for fully separation of each source.}

\end{table}

\subsection{Evaluation metrics}
\label{ssec:evaluation-metrics}
We use the signal-to-distortion (SDR) metric as our evaluation metric, which is also used by most audio source separation works. We directly use the API provided by the SiSec separation campaign 2018 \cite{14}, where evaluation is performed with tracks segmented to one second long excerpts. Meanwhile, medians of SDR are taken as evaluation metrics as well. As mentioned in \cite{5}, they consider medians to be more robust against outliers of extremely low SDR value caused by occasional near-silent parts in vocal tracks.

\subsection{Observations}
\label{ssec:observation}

Fig.2 takes the mean SDR statistic for each U-Net variant as the y-axis and the corresponding statistic for each J-Net variant as the x-axis. Therefore, each (J-Net,U-Net) pair can be visualized as one point.  Among all statistics, we find the mean SDR of output vocal has the strongest correlation. As we can observe the ceiling effect in the evaluation results of output accompaniments from Fig.3. High SDR value is easy to be achieved by any fully trained model. Therefore, the correlation of mean SDR statistics for separated accompaniment isn't in line with our expectations. As mentioned in \cite{5}, SDR is typically low when the separator output is quiet but not silent for a near-silent target. In comparison, high SDR out of the separation of vocal is much more difficult since there are lots of near-silent parts in the target vocal for models to overcome. From Fig.2, we can find a positive correlation among most pairs of models concerning the baseline pair (U\textsubscript{1}, J\textsubscript{1}). With (U\textsubscript{1}, J\textsubscript{1}) set as origin, we can observe that most points fall in the first and third quadrant. Those points fall in the first quadrant demonstrate that better performance with J-Net is an indication for better performance with corresponding U-Net, vice versa. We suppose that the correlation is stronger since the potential of each model variant is fully presented.

Meanwhile, some of our architecture variants perform better than the baseline model. Such as U2\textsubscript{2\_10}, U3\textsubscript{10}, U4\textsubscript{1} and U5 as  shown in Table 1. These results show the potential of each mean in Section 2 and possibly solve the relative hypothetical problems, such as semantic gaps or variation of scale in signals, in audio source separation task. From Table 3, we can find the effectiveness of the proposed progressive learning method. While the mean SDR of baseline model only gets the benefit from passing the second separation stage and the result starts to worsen from the third stage, we can observe the mean SDR of U4 keeps growing with more stages passed. However, the median is dropping at the same time. We suppose that is because the output vocals with worst SDR score are separated better after passing more stages, taking the output with the best SDR score as trade-off. With more stages, the near-silent parts where accompaniment remain within output are removed, thus, eliminating the outliers with worse SDR scores as mentioned in \cite{5}.

Finally, we discover a counter-intuitive fact that U-Net performs better with untrained decoder in comparison to random encoder. Surprisingly, the output result is even comparable to fully trained model U1 as shown in Table 2. We consider that shows the performance of U-Net is mostly determined by the skip connections from the first few convolution layers of encoder. To validate our hypothesis, we evaluate on models L\textsubscript{first-3} with skip connections only from the first three layers of encoder and the models L\textsubscript{last-3} with skip connections kept only for the last three layers of encoder. Notice that in both settings, the final output layer is still trained and the skip connection directly come from input signal is reserved. However, the results in Table 2 violate our hypothesis. We can find L\textsubscript{last-3} perform better and even outperform L which contains more skip connections. This observation leads us to rethink how skip connections are better added and how they function on top of U-Net.

\section{Conclusion}
\label{sec:conclusion}
In this paper, we propose to select network architecture or determine the effectiveness of some training tricks by performance evaluation on top of the model with a random encoder. Also, we demonstrate the proposed methods mentioned in Section 2 are beneficial for audio source separation task. Finally, we discover the potential of untrained decoder with aids from skip connections, not only outperform the model with random encoder but is comparable to the fully trained model. We leave validation of the positive correlation between models with random decoder and its fully trained counterparts as future work.

\section{Acknowledgement}
\label{sec:ack}
I would like to acknowledge the support provided by my friends, Ying-Tuan Hsu, Ming-Hao	Wen and Wei-En Lee during the preparation of this paper.

\vfill\pagebreak

\bibliographystyle{IEEEbib}
\bibliography{refs}

\end{document}